\documentclass[sn-standardnature]{sn-jnl}
\usepackage{commath}

\usepackage{xcolor}

\title{Confined vacuum resonances as artificial atoms with tunable lifetime}

\author[1]{\fnm{Rasa} \sur{Rejali}} 

\author[1]{\fnm{La\"{e}titia} \sur{Farinacci}}

\author[1]{\fnm{David} \sur{Coffey}}
\author[1]{\fnm{Rik} \sur{Broekhoven}}
\author[1]{\fnm{Jeremie} \sur{Gobeil}}
\author[1]{\fnm{Yaroslav M.} \sur{Blanter}}
\author*[1]{\fnm{Sander} \sur{Otte}} 

\email{a.f.otte@tudelft.nl}
\affil[1]{\orgdiv{Department of Quantum Nanoscience, Kavli Institute of Nanoscience}, \orgname{Delft University of Technology}, \orgaddress{\street{Lorentzweg 1}, \city{Delft}, \postcode{2628 CJ}, \country{The Netherlands}}}

\begin{document}

\abstract{Atomically engineered artificial lattices are a useful tool for simulating complex quantum phenomena, but have so far been limited to the study of Hamiltonians where electron-electron interactions do not play a role---but it's precisely the regime in which these interactions do matter where computational times lend simulations a critical advantage over numerical methods. Here, we propose a new platform for constructing artificial matter that relies on the confinement of field-emission resonances, a class of vacuum-localized discretized electronic states. We use atom manipulation of surface vacancies in a chlorine-terminated Cu(100) surface to reveal square patches of the underlying metal, thereby creating atomically-precise potential wells that host particle-in-a-box modes. By adjusting the shape and size of the confining potential, we can access states with different quantum numbers, making these patches attractive candidates as quantum dots or artificial atoms. We demonstrate that the lifetime of electrons in these engineered states can be extended and tuned through modification of the confining potential, either via atomic assembly or by changing the tip-sample distance. We also demonstrate control over a finite range of state-filling, a parameter which plays a key role in the evolution of quantum many-body states. We model the transport through the localized state to disentangle and quantify the lifetime-limiting processes, illustrating the critical dependency of the electron lifetime on the properties of the underlying bulk band structure. The interplay with the bulk bands also gives rise to negative differential resistance, opening possible avenues for engineering custom atomic-scale resonant tunnelling diodes, which exhibit similar current-voltage characteristics.}

\maketitle

Artificial lattices serve as quantum simulators for realizing and studying fundamental properties of real materials, with the advantage that the relevant interactions can be precisely controlled. While different experimental approaches, such as using ultra-cold atoms~\cite{BlochNatPhys2012}, optical lattices~\cite{BlochNatPhys2005,GrossScience2017}, or trapped ions~\cite{BlattNatPhys2012}, have been successfully implemented in the study of artificially constructed systems, atom manipulation casts the scanning tunneling microscope (STM) as a particularly appealing platform: the scanning probe framework uniquely allows for creating and characterizing the electronic properties of 2D artificial matter on the atomic scale~\cite{KhajetooriansNatRevPhy2019}. Typically, atomic impurities are patterned to construct a potential landscape that mimics a specific physical system, with the aim of studying model Hamiltonians. This approach has led to the realization of a wide range of novel states in, for instance, Dirac materials, like the Lieb lattice~\cite{DrostNaturePhys2017,SlotNatPhys2017} and artificial graphene~\cite{GomesNature2012,GardenierACSNano2020}, as well as higher order topological insulators~\cite{KempkesNatMat2019,FreeneyPRL2020}, among others~\cite{KempkesNatPhys2019, HudaNPJ2020,HudaPRR2020,GirovskySciPost2017,CollinsNatComm2017}. These studies offer rare insight into the parameters that govern the electronic behaviour of these systems, but are restricted by the short electron lifetime of the constituent artificial atoms to the limiting case in which electron-electron interactions do not play a role. Additionally, short electron lifetimes limit the available energy resolution; the most popular STM approach so far, which relies on confining surface states, lacks flexibility in tuning this parameter~\cite{BraunPRL2002,JensenPRB2005,KliewerNJofPhys2001}. 

Here, we explore a new platform for realizing artificial lattices, based on confining field-emission resonances (FERs): a class of quantized electronic states localized in the vacuum, between the surface and the probe tip, that arise in the high bias regime, i.e. exceeding the sample work function. We show that confining potentials can be engineered to enable the study of states with different orbital character~\cite{SlotPRX2019,GardenierACSNano2020,FreeneySciPost2020}, with precise control over the energy and quantum numbers of the states. We study the electron lifetime of these states, and demonstrate that we can finely tune it---and consequently, to some extent, the average occupation---by adjusting the tip-height or patch dimensions. The ability to tune the lifetime and occupation of artificial atoms is a critical first step towards simulating many-body quantum states driven by electron-electron interactions. We also observe specific voltage-current characteristics, namely negative differential resistance, which are analogous to those of resonant tunneling diodes~\cite{Sun1998IEEE}, making the confined FERs also suitable to possible applications in creating customizable, atomic scale diodes.

\begin{figure}
    \centering
    \includegraphics[width=\linewidth]{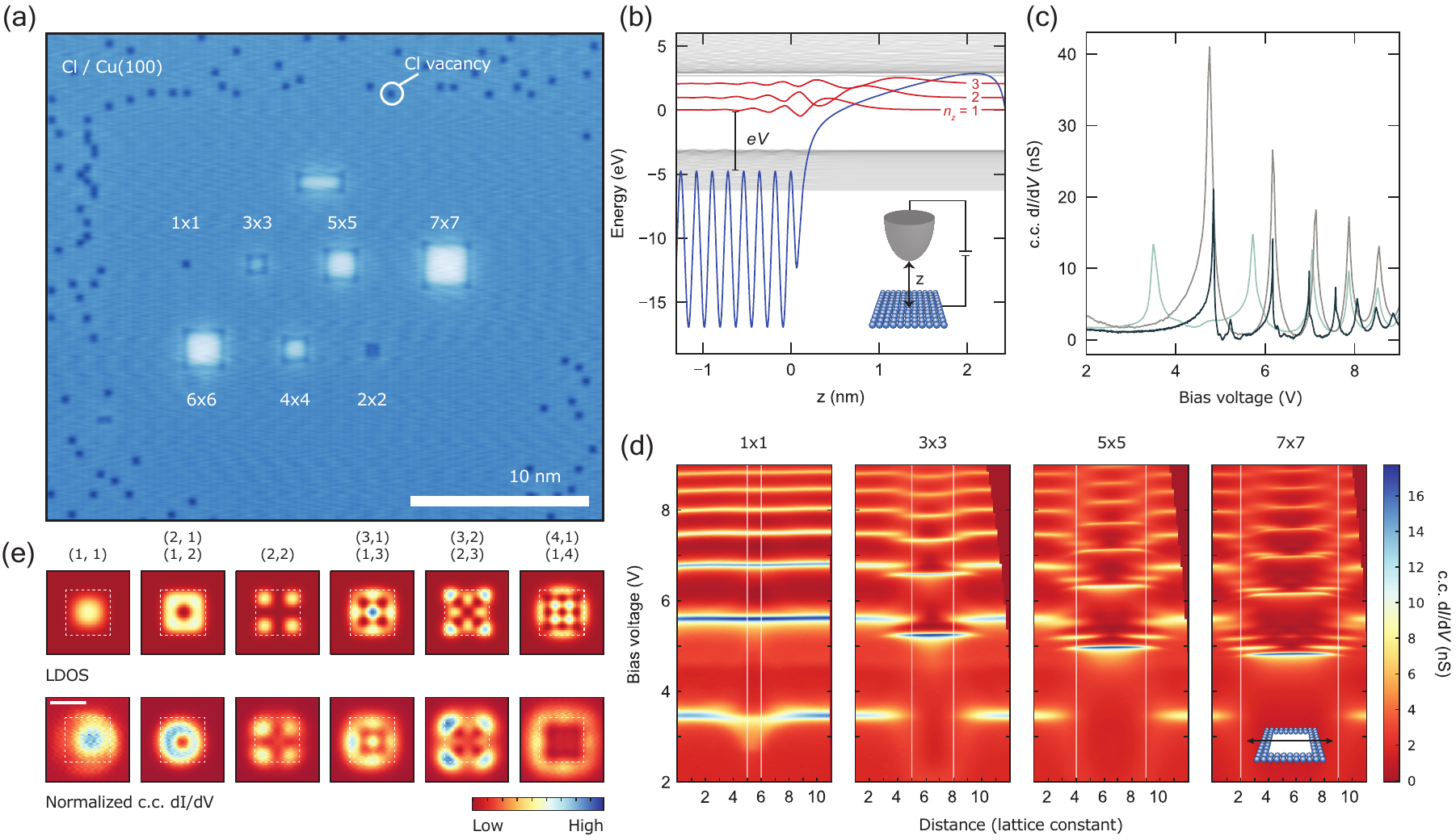}
    \caption{Confinement of field-emission resonances. (a) STM constant-current topography (600~mV, 300~pA) of square, atomically assembled patches of Cl vacancies, with sizes indicated in unit cells. (b) Potential landscape (blue) between sample (left) and tip (right) for a finite bias voltage $V$. Amongst the wave functions (grey) calculated for this potential, are the first three field-emission resonances (red). Inset: schematic of the tip-sample junction. (c) Constant-current differential conductance spectra acquired for bare Cu(100) (grey, 250~pA current setpoint), the chlorine monolayer (turquoise, 100~pA), and the center of the $7 \times 7$ patch (black, 100~pA). The first peak on the chlorine monolayer (3.5~V), being below the surface work function, corresponds to an image-potential state. (d) Stacked constant-current (100~pA) differential conductance spectra taken along a line crossing the center of each patch (shown in inset), with the corresponding patch size indicated (top). A correction is applied to the data to rectify the asymmetry of the tip electric field (see supplementary). White lines indicate the patch boundaries. (e) Calculated LDOS of the particle-in-a-box states ($\abs{\Psi}^2$), obtained using a finite well model (top row). Normalized~\cite{Rejali2022} constant-current (100~pA) differential conductance maps acquired for the $7 \times 7$ patch at the resonance energies of the first principal FER ($n_z=1$, $ (n_x, n_y) = (1,1)$) and the following sub-resonances. White squares delineate the spatial extent of the simulated potential well (top row) and the physical patch (bottom row). 
    }
    \label{fig:confinement}
\end{figure}

We use atom manipulation of single vacancies in the chlorine-terminated Cu(100) surface to engineer lateral confinement of field emission resonances. By coordinating chlorine vacancies--- which are easily manipulable and thus suited to large scale atomic assembly~\cite{GirovskySciPost2017,KalffNatureNano2016,DrostNaturePhys2017, HudaNPJ2020, HudaPRR2020}---adjacent to each other, we construct patches of exposed copper, surrounded by areas of homogeneous, monolayer chlorine coverage (figure~\ref{fig:confinement}a). The bare and chlorinated Cu(100) surfaces host FERs at bias voltages exceeding the local work function, at $4.6$~V~\cite{GartlandPRL1972} and $ 5.7$~V~\cite{WestphalSurfSci1983}, respectively. These resonances can be readily modelled with a one-dimensional potential in the out-of-plane direction (figure~\ref{fig:confinement}, see supplementary for details). The work function difference between the two surfaces results in a shift in the measured resonance energies (figure~\ref{fig:confinement}c), in accordance with previous studies~\cite{RuffieuxPRL2009,JungPRL1995,PivettaPRB2005, PloigtPRB2007}. 

Spectroscopy acquired at the center of the $7 \times 7$ patch (dimensions defined in unit cells of the chlorine lattice) exhibits additional resonances, in comparison to the bare and chlorinated Cu(100) surfaces (figure~\ref{fig:confinement}c). As shown in figure~\ref{fig:confinement}d, these additional resonances belong to a series of sub-resonances following each primary FER, and can in fact be resolved for each primary FER, up to and including the the fourth primary resonance. We use the principal quantum number $n_z$ to describe the primary FERs. Note that the additional modes are only observed above the energy of the first resonance on bare Cu(100) (figure~\ref{fig:confinement}c). The full in-plane structure of the confined modes for the larger patches is best visualised by differential conductance maps taken at voltages corresponding to the sub-resonances of the first FER on the $7 \times 7$ patch, as shown in figure~\ref{fig:confinement}e.  The observed states can be recognised as two-dimensional particle-in-a-box modes, with quantum numbers $n_x$ and $n_y$, and can be accurately reproduced by the eigenstates of a finite potential well (see supplementary for details). Similar to previous works~\cite{FreeneySciPost2020}, the nodal patterns of the first three modes are analogous to the orbitals of an two-dimensional atom, with the first state corresponding to the $s$-like state, and the second to the $p$-like, and subsequently the $d$-like state. Finally, we note that the energy of the FERs depends on the patch size: as the patch size is increased, the FER energy shifts down, tending toward the limit of bare Cu(100). All in all, the assembled patches can be seen as atomically precise potential wells, wherein the energy, spacing, and order of the states can be tuned by adjusting the shape and size of the confining potential.  We note that the single vacancy~\cite{GirovskySciPost2017,DrostNaturePhys2017,HudaNPJ2020,HudaPRR2020} stands out as an exception, as the necessary change in the local work function cannot take place on such small length-scales: as such, the vacancy acts as a scattering center, rather than a confinement potential.

\begin{figure}
    \centering
    \includegraphics[width=\linewidth]{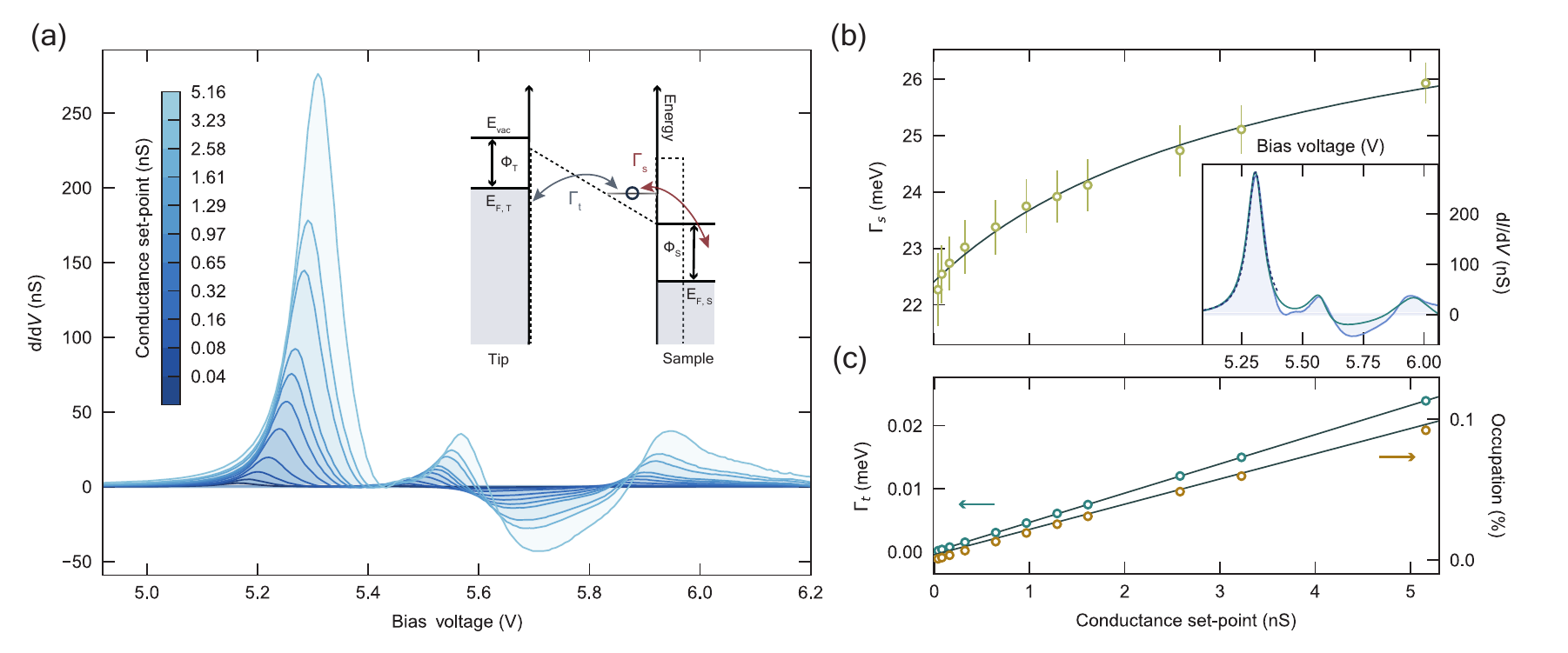}
    \caption{Extracting tip and sample decay rates. (a) Constant-height differential conductance spectra obtained at the center of the $5\times 5$ patch for a range of conductance setpoints ((250~pA $\to$ 32~nA), 6.2~V). Inset: schematic of the double-barrier potential (dotted line) implemented in the rate equations, with the decay rates to the tip and sample, $\Gamma_t$ and $\Gamma_s$, indicated. (b) Inset: constant-height differential conductance (light blue, shaded) acquired at the center of the $5\times 5$ patch (32~nA, 6.2~V). Calculated $\mathrm{d}/I\mathrm{d}V$ using a resonant tunneling model for a single level (navy, dotted line) or several, independent levels (green solid line). (b), (c) Sample (b, yellow circles) and tip (c, green circles) decay rates extracted for the first principal resonance as a function of conductance setpoint, fitted (solid grey line) to an inverse natural logarithm and a line, respectively. The tip decay rate is evaluated at the energy of the peak of the first principal field emission resonance. (c) Average occupation versus conductance setpoint (orange circles), and the corresponding linear fit (solid grey line). }
    \label{fig:5by5}
\end{figure}

In order to characterize the electron lifetime, we consider the transport through these confined states: two electron baths, one on the tip side and another on the sample side, act as decoherent sources, the contributions of which we can disentangle by investigating the evolution of the differential conductance spectra as a function of conductance setpoint, as shown in figure~\ref{fig:5by5}a. With increasing conductance setpoint, we observe a slight shift in the energy of the FERs, which is explained by the increased out-of-plane confinement (figure~\ref{fig:confinement}b), as well as the appearance of negative differential resistance (NDR). The appearance of NDR at high conductance setpoints gives us qualitative insight into the coupling of the resonances with the substrate and tip. 

We consider a transport model describing the resonant tunneling of independent electrons
from (to) the tip and sample through a level localized between the two potential barriers (2a, inset, see supplementary for more detail). In this framework, the current through a single resonance is given by:
\begin{equation}
   I_i = \frac{2G_Q\hbar}{e}\frac{\Gamma_{t}^i (z, V) \Gamma_{s}^i (z, V)}{\Gamma_{t}^i (z, V) + \Gamma_{s}^i (z, V)}\left( \frac{\pi}{2} + \tan^{-1}\left({\frac{2(eV - E_i(z,V))}{\hbar (\Gamma_{t}^i (z, V) + \Gamma_{s}^i (z, V))}}\right) \right),
    \label{eq:didv}
\end{equation}
where the quantum of conductance is $G_Q=e^2/(\pi\hslash)$, $\Gamma_{t}^i$ and $\Gamma_{s}^i$ are, respectively, the tip and sample decay rates for the $i$th resonance, and $E_i$ its energy, whose shift with bias voltage we will initially neglect for simplicity. In general, the tip and sample decay rates are both distance and voltage dependent. For the former, this dependence is derived by considering the transmission through the tunnel barrier. The sample decay rate, however, encapsulates an effective barrier that depends on the surface band-structure, and the relationship between $\Gamma_s$ and $V$ is non-trivial; we approximate this dependence as either constant or linear, depending on the width of the voltage window we consider. The differential conductance, in turn, can be obtained by differentiating the current with respect to voltage, and contains terms that scale with the derivatives of the decay rates and the energy of the resonance (see supplementary for the full expression).

We can gain quantitative insight into the tip and sample decay rates by focusing strictly on the first principal FER (figure~\ref{fig:5by5}b inset): this allows us to drastically reduce the number of free variables to a single resonance, and consequently to meaningfully account for the effects of the changing level $E_0$; additionally, we simplify $\Gamma_s(V)$ to a constant in the narrow voltage range around the resonance. By fitting the measured differential conductance at each conductance setpoint to our model, we can extract a value for the tip and sample decay rates as function of conductance setpoint (figure~\ref{fig:5by5}b and c).

In figure~\ref{fig:5by5}b, we see that $\Gamma_s$ increases with conductance setpoint, which can be related to the FER wave function: in general, decay to the bulk is governed by the overlap of the vacuum-localized state to the substrate, which is in turn determined via the penetration of the state into the bulk, the evanescent tail of the bulk states into the vacuum, and the diminished electronic screening in the area between the surface and the vacuum~\cite{BertholdPRL2002, EcheniqueChemPhys2000}. Bringing the tip closer causes a redistribution of the weight of the wave function toward the surface, rendering the scattering channels to the bulk more efficient~\cite{CrampinPRL2005_2}, leading to an increase in $\Gamma_s$. More precisely, we consider that the sample decay rate should scale linearly with the wave function overlap of the FER with the sample~\cite{BertholdPRL2002}, and for simplicity we assume its increase to be inversely proportional to the tip-sample distance. Given the exponential dependence of current with distance, we thus expect an inverse logarithmic dependence of the sample decay rate on the conductance setpoint. The fit in figure~\ref{fig:5by5}b shows this simple relation describes the change in $\Gamma_s$ appropriately.

The evolution of the tip decay rate with conductance setpoint is straightforward: this rate should scale exponentially with the tip-sample distance, meaning it should be linear with the conductance setpoint and intercept with the origin, as we see in figure~\ref{fig:5by5}c. Importantly, the changes in the decay rates impact the overall occupation of the state. The occupation is determined by the ratio of the tip decay rate to total decay rate $\Gamma_s  + \Gamma_t$, meaning that the occupation of the state can be tuned via the tip-height, as shown in figure~\ref{fig:5by5}c: the occupation linearly increases with the conductance setpoint. In effect, this means that the competing factors determining the time-average occupation---the rate of tunneling electrons versus the increase in the lifetime-limiting rate, $\Gamma_s$---results in the state filling increasing as the tip is brought closer.

We now extend our scope to account for transport through the higher energy states---around 5.6~V ($n_z =1$, $(n_x, n_y) = (2,1), (1,2)$) and 6~V ($(3,1), (1,3)$), respectively. To do so, we assume the resonances are independent, i.e. that the total current is determined by the sum of the currents $I_i$ through each resonance; additionally, we explicitly account for the voltage-dependence of $\Gamma_s(V)$ as linear to first approximation. As seen in figure~\ref{fig:5by5}b (inset), our model successfully reproduces the key features of the measured differential conductance over the entire voltage range, with, in particular, the presence of NDR between $\sim$ 5.6 to 6~V. In this window, we find $\rm{d} \Gamma_s/\rm{d} V < 0$. In fact, we find it is necessary  to have a decreasing sample decay rate with increasing voltage to engender NDR, indicating once again that the decay path to the sample crucially depends on the electronic wave function of the FER.

While the decay rates can be tuned by changing the out-of-plane confinement of the wave function, the in-plane confinement plays the dominant role in setting an upper bound on the lifetime. Typically, field-emission resonances are delocalized (Bloch-like) in the directions parallel to the surface and thus form bands~\cite{WalhPRL2003}. In that case, the electron lifetime is affected by interband scattering, wherein the excited electron escapes into the metal (sample or tip), or scatters with an electron in a different band; and intraband scattering, in which case the electron changes velocity~\cite{EcheniqueSurfSciReports2004,BertholdPRL2002}. We can expect the introduction of lateral localization to affect decay through these channels in two opposing ways: the increased confinement causes the bands to split into quantized states, strongly attenuating intraband decay, while the simultaneous broadening of the $k$-space distribution increases the available interband decay paths to the bulk. We assess the degree to which the in-plane confinement precisely affects the lifetime by investigating the transport characteristics of different sized patches.

\begin{figure}
    \centering
    \includegraphics[width=\linewidth]{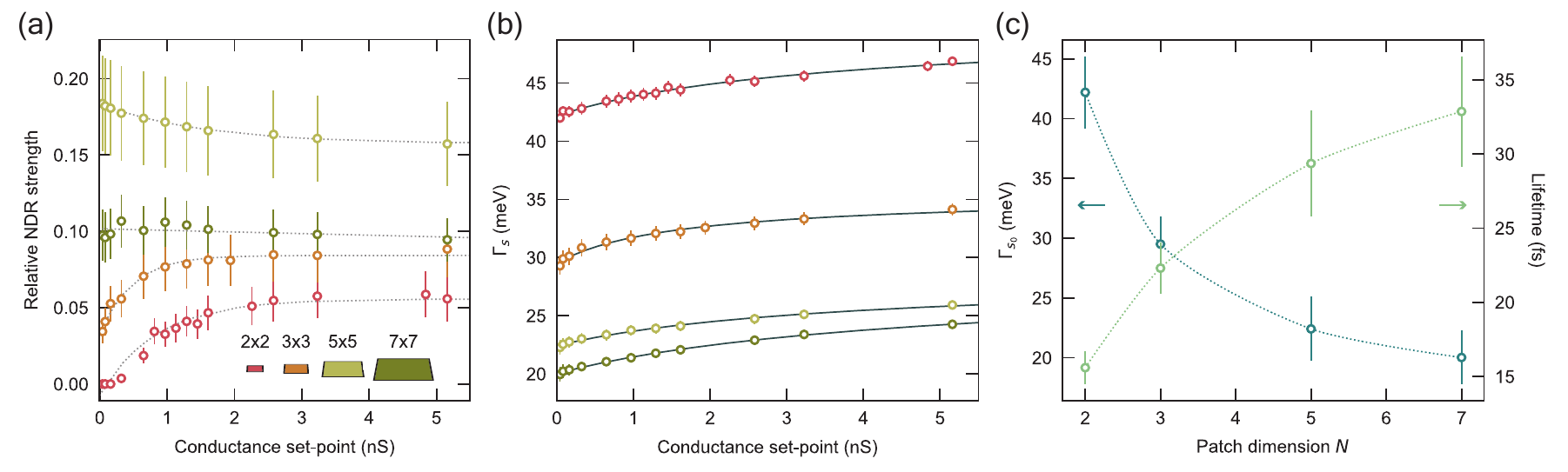}
    \caption{Tuning of the lifetime. (a) Relative strength of the negative differential resistance as a function of conductance setpoint, for patches of various size. Colors correspond to illustration in the inset. Dotted lines are guides to the eye. (b) Conductance-dependence of the sample decay rate for corresponding patch sizes, fitted to an inverse natural logarithmic function (grey solid lines). (c) Extrapolated value of the sample decay rate for zero setpoint conductance (blue circles), and the corresponding lifetime (green circles) for each patch size. Dotted lines are guides to eye. }
    \label{fig:lifetimes}
\end{figure}

Performing the same conductance-dependent measurements (see supplementary), we see a marked change in the relative strength of the NDR based on the dimensions of the confining patch, as shown in figure~\ref{fig:lifetimes}a. The relative NDR strength, which we define as the ratio of negative area to the total area under the differential conductance spectrum, stays fairly constant as a function of conductance setpoint for patches of larger size, such as the $7\times7$ and $5\times 5$. In contrast, the smallest patch ($2 \times 2$) does not exhibit any NDR at low conductance setpoints; at a conductance setpoint of $\sim$0.5~nS, the relative NDR strength becomes non-zero and monotonically increases thereafter. The same general trend holds for the $3\times 3$: exponentially increasing NDR strength with increasing conductance setpoint. In fact, the NDR is directly related to the change in the sample decay rate as a function of voltage, and we can see this variance in $\Gamma_s$ in the strength and conductance-dependent behaviour of the NDR for the different patches.

As before, to quantify the change in the sample decay rate, we extract $\Gamma_s$  by fitting equation~(\ref{eq:didv}) to the first principal FER of each patch, for a discrete range of conductance setpoints (figure~\ref{fig:lifetimes}b). We see that both the magnitude of the sample decay rate, and its rate of change over this conductance setpoint range, vary according to patch size. The electrons localized above the smallest patch experience the largest sample decay rates, meaning scattering to the bulk becomes more efficient due to the increased spatial confinement. 

The lifetime of these localized electrons, $\tau$, is determined by the tip and sample decay rates, such that $\tau^{-1} = \Gamma_s^{-1} + \Gamma_t^{-1}$. The tip contribution exponentially tends to zero as a function of the tip-sample distance, meaning the intrinsic lifetime (at zero conductance setpoint, namely when the tip is infinitely far away) is determined by the sample decay rate at zero conductance. Approximating the lifetime by the linewidth of the resonance is not valid here, as the potential in the out-of-plane direction changes as we perform spectroscopy, leading to a changing resonance energy as a function of the applied voltage that artificially broadens the peak.

As shown in figure~\ref{fig:lifetimes}c, the extracted lifetimes monotonically increase as a function of patch size up to $N=7$, the maximum patch dimension studied in this work. Notably, the lifetime for the confined states is roughly 2-4 times longer than the lifetime of the first resonance on bare Cu(100), extracted using the same method and in fair agreement with previously reported values (see supplementary). This also indicates that there must be a patch size with an optimally long lifetime, after which $\tau$ begins decreasing with patch size, tending toward the freely-propagating Cu(100) limit. Indeed, the degree to which the confinement prohibits the different decay paths at play is ultimately a delicate balance: the smaller the patch, the fewer states available for scattering between different resonances, but the larger the $k$-space overlap with the bulk states. Notably, the lifetime-limiting rate for all the patches shown here is $\Gamma_s$, which in our case is approximately three orders of magnitude larger than the tip decay rate $\Gamma_t$ (figure~\ref{fig:5by5}c).

\begin{figure}
    \centering
    \includegraphics[width=\linewidth]{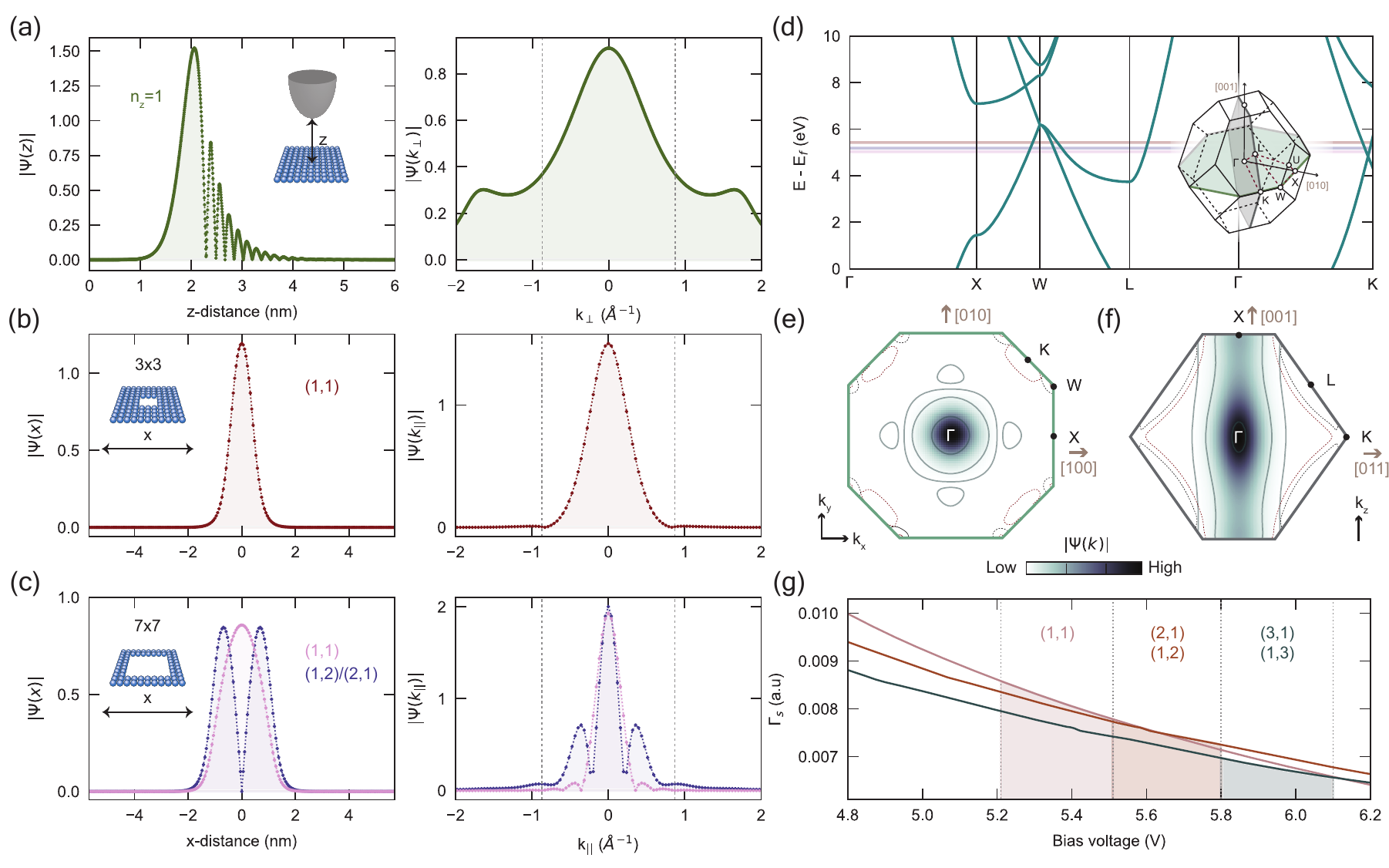}
    \caption{Distribution in $k$-space. (a) Calculated out-of-plane component of the real space wave function $\abs{\psi(z)}$ for the first principal FER $n_z = 0$ (left), and the corresponding Fourier transform (right), at tip-sample distance $z=2.4$~nm. (b), (c) Calculated in-plane component of the real space wave function (left) for the (b) $3\times 3$ (c) and $7\times 7$ patches, showing the  first $(n_x, n_y) = (1,1)$ ((b), red; (c), pink) and second $(1,2), (2,1)$ ((c), purple) modes, with the corresponding Fourier transforms (right). Dotted lines indicate $\pm \pi/a$ bounds. (d) Bulk band structure of Cu along high symmetry lines, with the experimental resonance energy of the $(1,1)$ state for the $3\times 3$ (red) and $7\times 7$ (pink), as well as the  $(1,2)/(2,1)$ state of the latter (purple), denoted by solid lines. Inset: schematic of the first Brillouin zone of Cu. (e, f) Intensity of the $3\times3$ wave function in \textit{k}-space across Brillouin zone slices indicated in inset of (d). Solid contour lines delineate an order of magnitude change in the intensity. Corresponding DFT-calculated constant-energy isolines shown for bulk Cu bands, taken 5~V (black line) and 6~V (red line) above the Fermi level. (g) Calculated sample decay rate as a function of bias voltage, shown for the first three resonances probed in the center of a $5\times5$ patch, corresponding to the (1,1) (mauve line), (2,1)/(1,2) (brown line), and (3,1)/(1,3) (grey line) modes. The shaded areas correspond to the voltage range in which the respective modes are typically measured, delineating $\Gamma_s$ in that range.}
    \label{fig:ffts}
\end{figure}

To better determine the role of the in- and out-of-plane confinement on the lifetime, we investigate the spectral weight of the localized resonances in $k$-space and compare this to the bulk band structure of copper. We calculate the wave function  $\Psi$, which we assume to be separable, in the directions parallel and perpendicular to the (100) direction to obtain the corresponding $k$-space distribution. First, we consider the out-of-plane direction, where the confinement is set by the tip-sample distance and the applied voltage. We restrict our focus to the calculated wave-function, $\Psi(z)$, for the first principal FER and the resulting Fourier transform, $\Psi(k_{\perp})$, shown in figure~\ref{fig:ffts}a. The $k_{\perp}$ values with a significant spectral weight span the entirety of the first Brillouin zone (BZ) ($\pm 1.75$~\AA). 

Along the in-plane directions, we consider the wave functions $\Psi(x)$ and $\Psi(y)$ corresponding to the first ($(1,1)$) particle-in-a-box mode for the $3\times 3$ and $7\times7$ patches (figure~\ref{fig:ffts}b and c). As expected, the $k_{\parallel}$-space distribution widens as the patch size decreases. Furthermore, as shown in figure~\ref{fig:ffts}c, this broadening also takes place when the quantum numbers $(n_x, n_y)$ of the in-plane mode increase. This is due to changes in the apparent barrier height: compare, for instance, the first and second particle-in-a-box modes---since the latter lies at higher energy than the former, it experiences a shallower finite well. Such considerations allow us to visualize how the factors considered so far---such as the tip-sample distance, the lateral extent of the patch, the apparent height of the in-plane barrier---impact the distribution of the state in $k$-space, and consequently its overlap with the bulk states.

%To better illustrate this, we look at cross-sections of the full $k$-space distribution of the wave function along different high-symmetry lines of the first Brillouin zone (BZ) (figure~\ref{fig:ffts}e-h), and the corresponding bulk band structure of bulk copper~\cite{BurdickPhysRev1963} (figure~\ref{fig:ffts}d). The horizontal cross-section of the BZ passes through the K, W, and X high symmetry points; the intensity and extent of the full $k$-space wave function, $\Psi(k)$, along this slice is shown in figure~\ref{fig:ffts}f. We also consider the vertical (figure~\ref{fig:ffts}g) and diagonal ((figure~\ref{fig:ffts}h) cross-section to be able to directly compare the extent of $\Psi(k)$ along every high symmetry shown in the band structure. In the energy range of interest (i.e. the experimental energies of the particle-in-a-box modes), denoted by solid lines in figure~\ref{fig:ffts}d, we cross several bulk bands across the high symmetry lines ($\rm{X} \to \rm{W}$, $\rm{W} \to \rm{L}$, $\rm{L} \to \Gamma$, $\Gamma \to \rm{K}$). 

To better illustrate this, we consider the band structure of bulk copper along the high symmetry lines~\cite{BurdickPhysRev1963}, specifically at the experimental energies of the particle-in-a-box modes (figure~\ref{fig:ffts}d). The lifetime of the confined electrons depends directly on, and is limited by, the number of bulk states available for direct tunneling---the more bands we cross at the energy of the resonance, with $k$-values falling within $\Psi(k)$, the shorter the lifetime to first order. In this energy range, we cross several bulk bands along the high symmetry lines ($\rm{X} \to \rm{W}$, $\rm{W} \to \rm{L}$, $\rm{L} \to \Gamma$, $\Gamma \to \rm{K}$); however, the efficiency of these decay paths is scaled by the spectral weight of $\Psi(k)$ at the crossing points. In other words, the efficiency of the decay paths is scaled by the probability of having an electron with the right momentum for direct tunneling into that bulk state.

Accordingly, in figures~\ref{fig:ffts}e and f, we consider the intensity of the $k$-space wave function along various cross-sections of the first BZ (figure~\ref{fig:ffts}d, inset). Interestingly, the highest spectral weight is along the $\Gamma \to \rm{X}$ direction---across both the lateral (figure~\ref{fig:ffts}e) and vertical (figure~\ref{fig:ffts}f) cross-sections---relative to the other high symmetry lines; however, this direction does not present any band crossing along the high symmetry lines at the energy of the resonances. In fact, $\Psi(k)$ carries little, if any, spectral weight along the other directions where it does cross the bulk bands. This is illustrated in figures~\ref{fig:ffts}e and f, where we see that $\Psi(k)$ has practically zero intensity along the energy isosurfaces (at 5~V and 6~V) of bulk Cu, calculated using density functional theory (DFT) (see supplementary for details). This is quite remarkable: although the lateral confinement of the states introduces direct tunneling paths to the bulk that are not present for the laterally freely-propagating case, we can consider the contribution to be minimal in this case. Additionally, the added confinement acts to largely hinder the role of intraband inelastic scattering, as the available states for scattering are substantially reduced: the FERs no longer form bands, but are rather quantized and well-separated in energy, according to the physical dimensions of the patch. These two effects ultimately amount to a considerable enhancement of the lifetime of the confined states.

These considerations also shed light on the dependence of the sample decay rate $\Gamma_s$ with bias voltage---which, as we previously found, is critical in engendering NDR. Namely: with increasing voltage, the localized resonance is pushed to higher energies, causing a shift in the crossing points with the bulk bands. In turn, this shift translates into the decay channels being scaled by a slightly different spectral weight. To illustrate this effect, we can consider the crossing along the $\Gamma \to \rm{K}$ direction: as the bias increases, the FER shifts up in energy, meaning that the crossing point for the lower band moves away from the $\Gamma$ point, closer to the $\rm{K}$ point. Figures~\ref{fig:ffts}e and f show that this shift is accompanied by a decrease in the spectral weight of $\Psi(k)$, meaning the total overlap between the localized state and the bulk bands decreases. The emergence of the upper band around $\sim$4.5V, however, further complicates the picture, illustrating that the overall rate of change of the decay rate is hard to estimate. However, by qualitatively considering the evolution of the $k$-space overlap, we can already grasp the complexity of the dependence of $\Gamma_s$ on the bias voltage.

To get a quantitative estimate of the change in the sample decay rate, we calculate the weighted $k$-space wave function overlap for each DFT-calculated crossing point throughout the entire BZ, and relate that to a dimensionless sample-decay rate via Fermi's golden rule (figure~\ref{fig:ffts}g, see supplementary for details). For this, we consider the calculated $k$-space wave function of the $5\times 5$ patch for the first ($1,1$), second ($2,1$), ($1,2$), and fourth ($3,1$), ($1,3$) particle-in-a-box modes---the only states with non-zero intensity at the center of the patch (see figures~\ref{fig:confinement}d and ~\ref{fig:5by5}a). As shown in figure~\ref{fig:ffts}g, we see that the calculated sample decay rate for all three states monotonically decreases, i.e. that the overlap of $\Psi(k)$ with the bulk bands decreases with increasing voltage, so that $\rm{d} \Gamma_s/\rm{d} V$ is negative---the ratio of this rate of change to the intercept is in good agreement with our quantitative results from the double barrier model (figure~\ref{fig:5by5}). The sample decay rate associated with each state is strictly only applicable in the voltage range in which that state is measured, roughly delineated in figure~\ref{fig:ffts}g by the shaded areas. All in all, we can confidently attribute the NDR to the effects of the bulk band structure. Additionally, we should also note that the NDR is consistently observed with different tips, and is not observed for laterally propagating FERs (see supplementary)~\cite{PascualPRB2007,StepanowPRB2011}, which do not have direct tunneling paths to the bulk available to them.

By laterally confining field-emission resonances through atomic assembly of single chlorine vacancies, we present a new platform for creating artificial atoms. We demonstrate control over the lifetime and occupation of these artificial atoms by adjusting the confining potential, implemented via modification of the tip-sample distance or the lateral dimensions of the patch. The ability to tune the occupation is a key parameter of control in the study of quantum many-body states that evolve as a function of the state filling. We show that the lifetime of field-emission resonances, unlike that of surface states, can be prolonged via lateral confinement, up to nearly four times the freely-propagating case. This extension of the lifetime enhances the available energy resolution, and, in conjunction with control over the state filling, is a first step towards studying electron-electron interactions with artificial lattices. Further prolonging the lifetime to approach a state occupation of 1 for reasonable setpoint currents can be pursued via several avenues: such as finding an underlying bulk crystal that hosts FER bands closer to the Fermi energy, or one that is semi-conducting or even insulating. These considerations make confined vacuum resonances a promising platform for creating and studying artificial lattices.

\section*{Methods}

Sample preparation and experimentation were performed in ultrahigh vacuum systems with a base pressure of $10^{-10}$~mbar (Unisoku USM1300s, SPECS Joule-Thompson-SPM). The Cu(100) crystal was cleaned via repeated cycles of argon sputter at 1~kV and annealing to 600$^\circ$~C. The chlorinated copper surface was prepared by thermal evaporation (2-3 minutes) of anhydrous CuCl$_2$ powder heated to 300$^\circ$~C onto a warm Cu(100) crystal. The crystal was heated to 150$^\circ$~C for $\sim 10$~minutes before and after deposition~\cite{KalffNatureNano2016}. The coverage and sample quality were verified via LEED (where possible) and STM. Atom manipulation of chlorine vacancies was implemented using a procedure previously outlined~\cite{KalffNatureNano2016}. Differential conductance measurements were performed using standard lock-in detection techniques.

\section{Data availability}
All data presented in this work are publicly available with identifier (DOI) 10.5281/zenodo.6473090

\section*{Acknowledgements} 
The authors thank the Dutch Research Council (NWO) and the European Research Council (ERC Starting Grant 676895 “SPINCAD”). The authors would like to thank Ingmar Swart, Anton Akhmerov, and Michael Wimmer for helpful discussions and insights, as well as Artem Pulkin for guidance with the DFT calculations.

\bibliography{main}

\begin{thebibliography}{10}
\expandafter\ifx\csname url\endcsname\relax
  \def\url#1{\burl{#1}}\fi
\expandafter\ifx\csname urlprefix\endcsname\relax\def\urlprefix{URL }\fi
\providecommand{\bibinfo}[2]{#2}
\providecommand{\eprint}[2][]{\url{#2}}
\providecommand{\doi}[1]{\url{https://doi.org/#1}}
\bibcommenthead

\bibitem{BlochNatPhys2012}
\bibinfo{author}{Bloch, I.}, \bibinfo{author}{Dalibard, J.} \&
  \bibinfo{author}{Nascimbène, S.}
\newblock \bibinfo{title}{Quantum simulations with ultracold quantum gases}.
\newblock \emph{\bibinfo{journal}{Nature Physics}}
  \textbf{\bibinfo{volume}{8}}~(4), \bibinfo{pages}{267--276}
  (\bibinfo{year}{2012}).
\newblock \doi{10.1038/nphys2259} .

\bibitem{BlochNatPhys2005}
\bibinfo{author}{Bloch, I.}
\newblock \bibinfo{title}{Ultracold quantum gases in optical lattices}.
\newblock \emph{\bibinfo{journal}{Nature Physics}}
  \textbf{\bibinfo{volume}{1}}~(1), \bibinfo{pages}{23--30}
  (\bibinfo{year}{2005}).
\newblock \doi{10.1038/nphys138} .

\bibitem{GrossScience2017}
\bibinfo{author}{Gross, C.} \& \bibinfo{author}{Bloch, I.}
\newblock \bibinfo{title}{Quantum simulations with ultracold atoms in optical
  lattices}.
\newblock \emph{\bibinfo{journal}{Science}}
  \textbf{\bibinfo{volume}{357}}~(6355), \bibinfo{pages}{995--1001}
  (\bibinfo{year}{2017}).
\newblock \doi{10.1126/science.aal3837} .

\bibitem{BlattNatPhys2012}
\bibinfo{author}{Blatt, R.} \& \bibinfo{author}{Roos, C.~F.}
\newblock \bibinfo{title}{Quantum simulations with trapped ions}.
\newblock \emph{\bibinfo{journal}{Nature Physics}}
  \textbf{\bibinfo{volume}{8}}~(4), \bibinfo{pages}{277--284}
  (\bibinfo{year}{2012}).
\newblock \urlprefix\url{https://doi.org/10.1038/nphys2252} .

\bibitem{KhajetooriansNatRevPhy2019}
\bibinfo{author}{Khajetoorians, A.~A.}, \bibinfo{author}{Wegner, D.},
  \bibinfo{author}{Otte, A.~F.} \& \bibinfo{author}{Swart, I.}
\newblock \bibinfo{title}{Creating designer quantum states of matter
  atom-by-atom}.
\newblock \emph{\bibinfo{journal}{Nature Reviews Physics}}
  \textbf{\bibinfo{volume}{1}}~(12), \bibinfo{pages}{703--715}
  (\bibinfo{year}{2019}).
\newblock \urlprefix\url{https://doi.org/10.1038/s42254-019-0108-5} .

\bibitem{DrostNaturePhys2017}
\bibinfo{author}{Drost, R.}, \bibinfo{author}{Ojanen, T.},
  \bibinfo{author}{Harju, A.} \& \bibinfo{author}{Liljeroth, P.}
\newblock \bibinfo{title}{Topological states in engineered atomic lattices}.
\newblock \emph{\bibinfo{journal}{Nature Physics}}
  \textbf{\bibinfo{volume}{13}}~(7), \bibinfo{pages}{668--671}
  (\bibinfo{year}{2017}).
\newblock \doi{10.1038/nphys4080} .

\bibitem{SlotNatPhys2017}
\bibinfo{author}{Slot, M.~R.} \emph{et~al.}
\newblock \bibinfo{title}{Experimental realization and characterization of an
  electronic {L}ieb lattice}.
\newblock \emph{\bibinfo{journal}{Nature Physics}}
  \textbf{\bibinfo{volume}{13}}~(7), \bibinfo{pages}{672--676}
  (\bibinfo{year}{2017}).
\newblock \doi{10.1038/nphys4105} .

\bibitem{GomesNature2012}
\bibinfo{author}{Gomes, K.~K.}, \bibinfo{author}{Mar, W.}, \bibinfo{author}{Ko,
  W.}, \bibinfo{author}{Guinea, F.} \& \bibinfo{author}{Manoharan, H.~C.}
\newblock \bibinfo{title}{Designer {D}irac fermions and topological phases in
  molecular graphene}.
\newblock \emph{\bibinfo{journal}{Nature}}
  \textbf{\bibinfo{volume}{483}}~(7389), \bibinfo{pages}{306--310}
  (\bibinfo{year}{2012}).
\newblock \doi{10.1038/nature10941} .

\bibitem{GardenierACSNano2020}
\bibinfo{author}{Gardenier, T.~S.} \emph{et~al.}
\newblock \bibinfo{title}{p orbital flat band and {D}irac cone in the
  electronic honeycomb lattice}.
\newblock \emph{\bibinfo{journal}{ACS Nano}}
  \textbf{\bibinfo{volume}{14}}~(10), \bibinfo{pages}{13638--13644}
  (\bibinfo{year}{2020}).
\newblock \doi{10.1021/acsnano.0c05747} .

\bibitem{KempkesNatMat2019}
\bibinfo{author}{Kempkes, S.~N.} \emph{et~al.}
\newblock \bibinfo{title}{Robust zero-energy modes in an electronic
  higher-order topological insulator}.
\newblock \emph{\bibinfo{journal}{Nature Materials}}
  \textbf{\bibinfo{volume}{18}}~(12), \bibinfo{pages}{1292--1297}
  (\bibinfo{year}{2019}).
\newblock \doi{10.1038/s41567-018-0328-0} .

\bibitem{FreeneyPRL2020}
\bibinfo{author}{Freeney, S.~E.}, \bibinfo{author}{van~den Broeke, J.~J.},
  \bibinfo{author}{Harsveld van~der Veen, A. J.~J.}, \bibinfo{author}{Swart,
  I.} \& \bibinfo{author}{Morais~Smith, C.}
\newblock \bibinfo{title}{Edge-dependent topology in kekul\'e lattices}.
\newblock \emph{\bibinfo{journal}{Phys. Rev. Lett.}}
  \textbf{\bibinfo{volume}{124}}, \bibinfo{pages}{236404}
  (\bibinfo{year}{2020}).
\newblock \doi{10.1103/PhysRevLett.124.236404} .

\bibitem{KempkesNatPhys2019}
\bibinfo{author}{Kempkes, S.~N.} \emph{et~al.}
\newblock \bibinfo{title}{Design and characterization of electrons in a fractal
  geometry}.
\newblock \emph{\bibinfo{journal}{Nature Physics}}
  \textbf{\bibinfo{volume}{15}}~(2), \bibinfo{pages}{127--131}
  (\bibinfo{year}{2019}).
\newblock \doi{10.1038/s41567-018-0328-0} .

\bibitem{HudaNPJ2020}
\bibinfo{author}{Huda, M.~N.}, \bibinfo{author}{Kezilebieke, S.},
  \bibinfo{author}{Ojanen, T.}, \bibinfo{author}{Drost, R.} \&
  \bibinfo{author}{Liljeroth, P.}
\newblock \bibinfo{title}{Tuneable topological domain wall states in engineered
  atomic chains}.
\newblock \emph{\bibinfo{journal}{npj Quantum Materials}}
  \textbf{\bibinfo{volume}{5}}~(1), \bibinfo{pages}{17} (\bibinfo{year}{2020}).
\newblock \doi{10.1038/s41535-020-0219-3} .

\bibitem{HudaPRR2020}
\bibinfo{author}{Huda, M.~N.}, \bibinfo{author}{Kezilebieke, S.} \&
  \bibinfo{author}{Liljeroth, P.}
\newblock \bibinfo{title}{Designer flat bands in quasi-one-dimensional atomic
  lattices}.
\newblock \emph{\bibinfo{journal}{Phys. Rev. Research}}
  \textbf{\bibinfo{volume}{2}}, \bibinfo{pages}{043426} (\bibinfo{year}{2020}).
\newblock \doi{10.1103/PhysRevResearch.2.043426} .

\bibitem{GirovskySciPost2017}
\bibinfo{author}{Girovsky, J.} \emph{et~al.}
\newblock \bibinfo{title}{Emergence of quasiparticle {{Bloch}} states in
  artificial crystals crafted atom-by-atom}.
\newblock \emph{\bibinfo{journal}{SciPost Physics}}
  \textbf{\bibinfo{volume}{2}}~(3), \bibinfo{pages}{020}
  (\bibinfo{year}{2017}).
\newblock \doi{10.21468/SciPostPhys.2.3.020} .

\bibitem{CollinsNatComm2017}
\bibinfo{author}{Collins, L.~C.}, \bibinfo{author}{Witte, T.~G.},
  \bibinfo{author}{Silverman, R.}, \bibinfo{author}{Green, D.~B.} \&
  \bibinfo{author}{Gomes, K.~K.}
\newblock \bibinfo{title}{Imaging quasiperiodic electronic states in a
  synthetic penrose tiling}.
\newblock \emph{\bibinfo{journal}{Nature Communications}}
  \textbf{\bibinfo{volume}{8}}~(1), \bibinfo{pages}{15961}
  (\bibinfo{year}{2017}).
\newblock \doi{10.1038/ncomms15961} .

\bibitem{BraunPRL2002}
\bibinfo{author}{Braun, K.-F.} \& \bibinfo{author}{Rieder, K.-H.}
\newblock \bibinfo{title}{Engineering electronic lifetimes in artificial atomic
  structures}.
\newblock \emph{\bibinfo{journal}{Phys. Rev. Lett.}}
  \textbf{\bibinfo{volume}{88}}, \bibinfo{pages}{096801}
  (\bibinfo{year}{2002}).
\newblock \doi{10.1103/PhysRevLett.88.096801} .

\bibitem{JensenPRB2005}
\bibinfo{author}{Jensen, H.}, \bibinfo{author}{Kr\"oger, J.},
  \bibinfo{author}{Berndt, R.} \& \bibinfo{author}{Crampin, S.}
\newblock \bibinfo{title}{Electron dynamics in vacancy islands: Scanning
  tunneling spectroscopy on {A}g(111)}.
\newblock \emph{\bibinfo{journal}{Phys. Rev. B}} \textbf{\bibinfo{volume}{71}},
  \bibinfo{pages}{155417} (\bibinfo{year}{2005}).
\newblock \doi{10.1103/PhysRevB.71.155417} .

\bibitem{KliewerNJofPhys2001}
\bibinfo{author}{Kliewer, J.}, \bibinfo{author}{Berndt, R.} \&
  \bibinfo{author}{Crampin, S.}
\newblock \bibinfo{title}{Scanning tunnelling spectroscopy of electron
  resonators}.
\newblock \emph{\bibinfo{journal}{New Journal of Physics}}
  \textbf{\bibinfo{volume}{3}}, \bibinfo{pages}{22--22} (\bibinfo{year}{2001}).
\newblock \doi{10.1088/1367-2630/3/1/322} .

\bibitem{SlotPRX2019}
\bibinfo{author}{Slot, M.~R.} \emph{et~al.}
\newblock \bibinfo{title}{$p$-band engineering in artificial electronic
  lattices}.
\newblock \emph{\bibinfo{journal}{Phys. Rev. X}} \textbf{\bibinfo{volume}{9}},
  \bibinfo{pages}{011009} (\bibinfo{year}{2019}).
\newblock \urlprefix\url{https://link.aps.org/doi/10.1103/PhysRevX.9.011009}.
\newblock \doi{10.1103/PhysRevX.9.011009} .

\bibitem{FreeneySciPost2020}
\bibinfo{author}{Freeney, S.}, \bibinfo{author}{Borman, S.},
  \bibinfo{author}{Harteveld, J.} \& \bibinfo{author}{Swart, I.}
\newblock \bibinfo{title}{{Coupling quantum corrals to form artificial
  molecules}}.
\newblock \emph{\bibinfo{journal}{SciPost Phys.}} \textbf{\bibinfo{volume}{9}},
  \bibinfo{pages}{85} (\bibinfo{year}{2020}).
\newblock \doi{10.21468/SciPostPhys.9.6.085} .

\bibitem{Sun1998IEEE}
\bibinfo{author}{Sun, J.~P.}, \bibinfo{author}{Haddad, G.},
  \bibinfo{author}{Mazumder, P.} \& \bibinfo{author}{Schulman, J.}
\newblock \bibinfo{title}{Resonant tunneling diodes: models and properties}.
\newblock \emph{\bibinfo{journal}{Proceedings of the IEEE}}
  \textbf{\bibinfo{volume}{86}}~(4), \bibinfo{pages}{641--660}
  (\bibinfo{year}{1998}).
\newblock \doi{10.1109/5.663541} .

\bibitem{KalffNatureNano2016}
\bibinfo{author}{Kalff, F.~E.} \emph{et~al.}
\newblock \bibinfo{title}{A kilobyte rewritable atomic memory}.
\newblock \emph{\bibinfo{journal}{Nature Nanotechnology}}
  \textbf{\bibinfo{volume}{11}}~(11), \bibinfo{pages}{926--929}
  (\bibinfo{year}{2016}).
\newblock \doi{10.1038/nnano.2016.131} .

\bibitem{GartlandPRL1972}
\bibinfo{author}{Gartland, P.~O.}, \bibinfo{author}{Berge, S.} \&
  \bibinfo{author}{Slagsvold, B.~J.}
\newblock \bibinfo{title}{Photoelectric work function of a copper single
  crystal for the (100), (110), (111), and (112) faces}.
\newblock \emph{\bibinfo{journal}{Phys. Rev. Lett.}}
  \textbf{\bibinfo{volume}{28}}, \bibinfo{pages}{738--739}
  (\bibinfo{year}{1972}).
\newblock \urlprefix\url{https://link.aps.org/doi/10.1103/PhysRevLett.28.738}.
\newblock \doi{10.1103/PhysRevLett.28.738} .

\bibitem{WestphalSurfSci1983}
\bibinfo{author}{Westphal, D.} \& \bibinfo{author}{Goldmann, A.}
\newblock \bibinfo{title}{Chlorine adsorption on copper: {II}. photoemission
  from {C}u(001)c($2 \times 2$)-{C}l and {C}u(111)($\sqrt{3} \times
  \sqrt{3}$){R}30$^{\circ}$-{C}l}.
\newblock \emph{\bibinfo{journal}{Surface Science}}
  \textbf{\bibinfo{volume}{131}}~(1), \bibinfo{pages}{113--138}
  (\bibinfo{year}{1983}).
\newblock
  \urlprefix\url{https://www.sciencedirect.com/science/article/pii/003960288390122X}.
\newblock \doi{https://doi.org/10.1016/0039-6028(83)90122-X} .

\bibitem{RuffieuxPRL2009}
\bibinfo{author}{Ruffieux, P.} \emph{et~al.}
\newblock \bibinfo{title}{Mapping the electronic surface potential of
  nanostructured surfaces}.
\newblock \emph{\bibinfo{journal}{Phys. Rev. Lett.}}
  \textbf{\bibinfo{volume}{102}}, \bibinfo{pages}{086807}
  (\bibinfo{year}{2009}).
\newblock \doi{10.1103/PhysRevLett.102.086807} .

\bibitem{JungPRL1995}
\bibinfo{author}{Jung, T.}, \bibinfo{author}{Mo, Y.~W.} \&
  \bibinfo{author}{Himpsel, F.~J.}
\newblock \bibinfo{title}{Identification of metals in scanning tunneling
  microscopy via image states}.
\newblock \emph{\bibinfo{journal}{Phys. Rev. Lett.}}
  \textbf{\bibinfo{volume}{74}}, \bibinfo{pages}{1641--1644}
  (\bibinfo{year}{1995}).
\newblock \doi{10.1103/PhysRevLett.74.1641} .

\bibitem{PivettaPRB2005}
\bibinfo{author}{Pivetta, M.}, \bibinfo{author}{Patthey, F. m.~c.},
  \bibinfo{author}{Stengel, M.}, \bibinfo{author}{Baldereschi, A.} \&
  \bibinfo{author}{Schneider, W.-D.}
\newblock \bibinfo{title}{Local work function {M}oir\'e pattern on ultrathin
  ionic films: {N}a{C}l on {A}g(100)}.
\newblock \emph{\bibinfo{journal}{Phys. Rev. B}} \textbf{\bibinfo{volume}{72}},
  \bibinfo{pages}{115404} (\bibinfo{year}{2005}).
\newblock \doi{10.1103/PhysRevB.72.115404} .

\bibitem{PloigtPRB2007}
\bibinfo{author}{Ploigt, H.-C.}, \bibinfo{author}{Brun, C.},
  \bibinfo{author}{Pivetta, M.}, \bibinfo{author}{Patthey, F. m.~c.} \&
  \bibinfo{author}{Schneider, W.-D.}
\newblock \bibinfo{title}{Local work function changes determined by field
  emission resonances: {N}a{C}l/{Ag}(100)}.
\newblock \emph{\bibinfo{journal}{Phys. Rev. B}} \textbf{\bibinfo{volume}{76}},
  \bibinfo{pages}{195404} (\bibinfo{year}{2007}) .

\bibitem{BertholdPRL2002}
\bibinfo{author}{Berthold, W.} \emph{et~al.}
\newblock \bibinfo{title}{Momentum-resolved lifetimes of image-potential states
  on {C}u(100)}.
\newblock \emph{\bibinfo{journal}{Phys. Rev. Lett.}}
  \textbf{\bibinfo{volume}{88}}, \bibinfo{pages}{056805}
  (\bibinfo{year}{2002}).
\newblock \doi{10.1103/PhysRevLett.88.056805} .

\bibitem{EcheniqueChemPhys2000}
\bibinfo{author}{Echenique, P.}, \bibinfo{author}{Pitarke, J.},
  \bibinfo{author}{Chulkov, E.} \& \bibinfo{author}{Rubio, A.}
\newblock \bibinfo{title}{Theory of inelastic lifetimes of low-energy electrons
  in metals}.
\newblock \emph{\bibinfo{journal}{Chemical Physics}}
  \textbf{\bibinfo{volume}{251}}~(1), \bibinfo{pages}{1--35}
  (\bibinfo{year}{2000}).
\newblock \doi{10.1016/s0301-0104(99)00313-4} .

\bibitem{CrampinPRL2005_2}
\bibinfo{author}{Crampin, S.}
\newblock \bibinfo{title}{Lifetimes of stark-shifted image states}.
\newblock \emph{\bibinfo{journal}{Phys. Rev. Lett.}}
  \textbf{\bibinfo{volume}{95}}, \bibinfo{pages}{046801}
  (\bibinfo{year}{2005}).
\newblock \doi{10.1103/PhysRevLett.95.046801} .

\bibitem{WalhPRL2003}
\bibinfo{author}{Wahl, P.}, \bibinfo{author}{Schneider, M.~A.},
  \bibinfo{author}{Diekh\"oner, L.}, \bibinfo{author}{Vogelgesang, R.} \&
  \bibinfo{author}{Kern, K.}
\newblock \bibinfo{title}{Quantum coherence of image-potential states}.
\newblock \emph{\bibinfo{journal}{Phys. Rev. Lett.}}
  \textbf{\bibinfo{volume}{91}}, \bibinfo{pages}{106802}
  (\bibinfo{year}{2003}).
\newblock \doi{10.1103/PhysRevLett.91.106802} .

\bibitem{EcheniqueSurfSciReports2004}
\bibinfo{author}{Echenique, P.} \emph{et~al.}
\newblock \bibinfo{title}{Decay of electronic excitations at metal surfaces}.
\newblock \emph{\bibinfo{journal}{Surface Science Reports}}
  \textbf{\bibinfo{volume}{52}}~(7), \bibinfo{pages}{219--317}
  (\bibinfo{year}{2004}).
\newblock \doi{10.1016/j.surfrep.2004.02.002} .

\bibitem{BurdickPhysRev1963}
\bibinfo{author}{Burdick, G.~A.}
\newblock \bibinfo{title}{Energy band structure of copper}.
\newblock \emph{\bibinfo{journal}{Phys. Rev.}} \textbf{\bibinfo{volume}{129}},
  \bibinfo{pages}{138--150} (\bibinfo{year}{1963}).
\newblock \doi{10.1103/PhysRev.129.138} .

\bibitem{PascualPRB2007}
\bibinfo{author}{Pascual, J.~I.} \emph{et~al.}
\newblock \bibinfo{title}{Role of the electric field in surface electron
  dynamics above the vacuum level}.
\newblock \emph{\bibinfo{journal}{Phys. Rev. B}} \textbf{\bibinfo{volume}{75}},
  \bibinfo{pages}{165326} (\bibinfo{year}{2007}).
\newblock \doi{10.1103/PhysRevB.75.165326} .

\bibitem{StepanowPRB2011}
\bibinfo{author}{Stepanow, S.} \emph{et~al.}
\newblock \bibinfo{title}{Localization, splitting, and mixing of field emission
  resonances induced by alkali metal clusters on {C}u(100)}.
\newblock \emph{\bibinfo{journal}{Phys. Rev. B}} \textbf{\bibinfo{volume}{83}},
  \bibinfo{pages}{115101} (\bibinfo{year}{2011}).
\newblock \doi{10.1103/PhysRevB.83.115101} .

\bibitem{Rejali2022}
\bibinfo{author}{Rejali, R.}, \bibinfo{author}{Farinacci, L.} \&
  \bibinfo{author}{Otte, S.}
\newblock \bibinfo{title}{Normalization procedure for obtaining the local
  density of states from high-bias scanning tunneling spectroscopy}.
\newblock \emph{\bibinfo{journal}{arXiv preprint}}  (\bibinfo{year}{2022}).
\newblock \doi{https://doi.org/10.48550/arXiv.2204.09929} .

\end{thebibliography}
\end{document}